\documentstyle[aps,tighten,preprint,epsf]{revtex}

\newlength{\len}

\draft
\begin{document}
\title{Spectral Statistics Beyond Random Matrix Theory.
}
\author{A.~V.~Andreev and B.~L.~Altshuler}
\address{
NECI, 4 Independence Way, Princeton, NJ\ 08540, USA,\\
Department of Physics, Massachusetts Institute of Technology, 77 
Massachusetts Avenue, Cambridge, MA\ 02139, USA
}

\maketitle

\begin{abstract}
Using a nonperturbative approach we examine the large frequency 
asymptotics of the two-point level density correlator 
in weakly disordered metallic grains. 
This allows us to study the behavior of the two-level 
structure factor close to the Heisenberg time. We find that the singularities 
(present for random matrix ensembles) are washed out in a grain with 
a finite conductance. The results are nonuniversal (they depend on the 
shape of the grain and on its conductance), though they suggest 
a generalization for any system with finite Heisenberg time. 
\end{abstract}

\narrowtext


A  great variety of physical systems are known to exhibit quantum 
chaos. The common examples are atomic nuclei, Rydberg atoms in a 
strong magnetic field, electrons in 
disordered metals, etc \cite{LesHouches89}. 
Chaotic behavior  manifests itself in the energy level 
statistics. It was a remarkable discovery of Wigner and Dyson, 
that these statistics in a particular system can be approximated 
by those of an ensemble of  random matrices (RM). Here we consider 
deviations from  the RM theory taking an ensemble of weakly disordered 
metallic grains with a finite conductance $g$ as an example.
The results seem to be extendible to general chaotic systems.

There are two characteristic energy scales associated with 
a particular system: a classical one $E_c$ and a quantum one. 
The quantum energy scale is the mean level spacing $\Delta$.
In a chaotic billiard, for example,  $E_c$ is set by the 
frequency of the shortest periodic orbit. Well developed chaotic
behavior can take place only if $E_c\gg \Delta$. 

In a disordered 
metallic grain the classical energy is the Thouless energy $E_c=D/L^2$, 
where $D$ is the diffusion constant, and $L$ is the system size. 
 For a weakly 
disordered grain the two scales are separated by 
the dimensionless conductance
$g=E_c/\Delta \gg 1$ \cite{Thouless}. 
For frequencies  $\omega \ll E_c$ the behavior of 
the system becomes universal (independent of particular parameters 
of the system ). 
In this regime in the zeroth approximation 
the level statistics  depend only  
on the symmetry of the system and are 
described by one of the RM ensembles:
unitary, orthogonal or symplectic \cite{Mehta91}.
 
\begin{figure}
  \begin{center}
    \epsfxsize=12cm \epsfbox{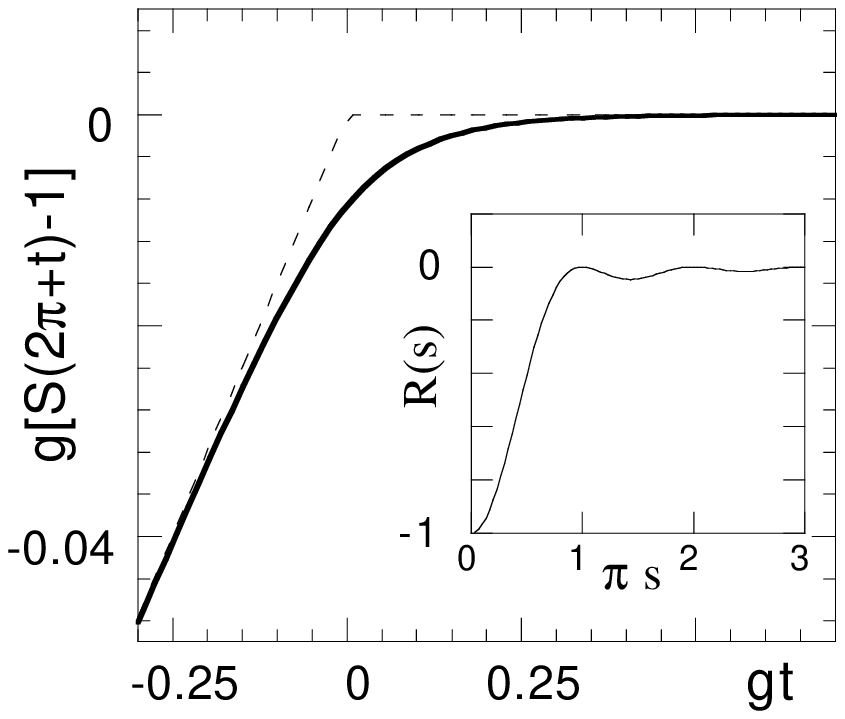}
  \end{center}
  \caption{Structure factor  
     in quasi-1D case for unitary symmetry (solid line) 
     and the universal structure factor (dashed line). 
     Insert: the two level correlator 
     as a function of level separation.}
  \label{fig:1}
\end{figure}

One of the conventional statistical spectral characteristics  is the   
two-point level density correlator 
\begin{equation}
K(\omega ,x)=
\langle \rho (\epsilon +\omega, \hat{H}+x\delta\hat{H} )
\rho (\epsilon , \hat{H})\rangle -\Delta^{-2}, 
\label{Kparam}
\end{equation}

\noindent where 
 $\hat{H}$ is the Hamiltonian of the system, 
$\delta\hat{H}$ is a perturbation, $x$ is the dimensionless 
perturbation strength and
 $\rho (\epsilon,\hat{H}+x\delta\hat{H})={\rm Tr}
\delta(\epsilon -\hat{H}-x\delta\hat{H})$ 
is the $x$-dependent density of states at 
energy $\epsilon$. It is convenient to introduce 
the  dimensionless frequency $s= \omega/\Delta$ and the 
dimensionless two-level correlator $R(s,x)=\Delta^2 K(\omega ,x)$. 
Dyson \cite{Dyson} determined $R(s,x=0)$ for RM.
For example, $ R(s ,o) $ in the unitary case equals to 
\begin{equation}
 R(s ,0)
        = \delta(s)-
\frac{1-\cos(2 \pi s)}{2(\pi s)^2}
        \label{Runivers}
\end{equation}
and is plotted in the insert in Fig.~\ref{fig:1}.

Perhaps the most striking signature of the Wigner-Dyson 
statistics is 
the rigidity of the energy spectrum \cite{Haake}.
Among the major consequences of this phenomenon are:
a) the probability to find two levels separated by $\omega \ll \Delta$ 
vanishes as $\omega \to 0$;
b) the level number variance in an energy strip of width $N\Delta$ 
is proportional to $\ln N$ rather than $N$; and  
c)  oscillations in the 
correlator $R(s , 0)$ in Eq.~(\ref{Runivers}) decay only algebraically. 

In the two level structure factor \cite{Prigodin94}
$S(\tau, x)=\int^{\infty}_{-\infty} ds \exp(i\tau s) R(s , x)$ 
the reduced fluctuations of the level number manifest themselves 
in the vanishing of  $S(\tau, 0)$ at 
$\tau=0$, and the algebraic decay of the oscillations in $R(s ,0)$ 
leads to the singularity in $S(\tau, 0)$ at the 
Heisenberg time $\tau=2\pi$. 
In the unitary case, e.~g.

\[ S(\tau, 0)={\rm min}\{|\tau|/(2\pi), 1\} 
.\]

\noindent At $\tau \ll 2\pi $  this Dyson result 
was obtained by Berry \cite{Berry85} for a generic chaotic 
system by  use of a 
semiclassical approximation. 
To the best of our knowledge nobody succeeded in analyzing the 
behavior of $S(\tau, 0)$ around $\tau=2\pi$ using this formalism.

Wigner-Dyson statistics become exact in the limit $g=E_c/\Delta \to \infty$.
We consider corrections to these statistics for finite $g$.
One of the better understood systems in this respect is a
weakly disordered metallic grain.
 For frequencies much smaller than $E_c$ the statistics are close 
to universal ones, the corrections being small as $(s/g)^2$
\cite{Kravtsov94}. At $s\gg 1$ the monotonic part 
of $R(s,x)$ can be 
obtained  perturbatively \cite{Altshuler85} 
\begin{equation}
  \label{Rpert}
R_{{\rm p}}(s,x)=
\Re\sum_{\mu}
\frac{1}{\alpha \pi^2(-is+ x^2 + \epsilon _\mu)^2},
\end{equation}
 
\noindent where $\epsilon_\mu$ are eigenvalues (in units of $\Delta$)
 of the 
diffusion equation in the grain,   $\alpha=2$ for the 
unitary ensemble and  $\alpha=1$ for the orthogonal and symplectic 
ensembles \cite{footnote}.
At this point we can define 
\begin{equation}
  \label{gdef}
  E_c=\epsilon_1 \Delta/\pi^2, {\qquad} g=\epsilon_1/\pi^2,
\end{equation}
where $\epsilon_1$ is the smallest nonzero eigenvalue. 
Perturbation theory allows one to
determine $S(\tau, 0)$ at small times $\tau \ll 1$. 
Since the oscillatory part of $R(s,x)$ is non-analytic in $1/s$ 
it can not be obtained perturbatively. 
 
In this Letter we 
obtain the leading $s \gg 1$ asymptotics of $R(s , x)$
 retaining the oscillatory terms 
and  monitor how the singularity in $S(\tau, 0)$ 
at the Heisenberg time  is modified by the finite conductance $g$. 
We make use of the  nonterturbative approach 
\cite{Efetov83} that is valid for arbitrary relation between 
$s$ and $g$. 
The oscillatory part $R_{\rm osc}(s,x)\equiv R(s,x)-R_{\rm p}(s,x)$ 
for the unitary (${\rm u}$), orthogonal (${\rm o}$) and 
 symplectic (${\rm s}$) cases equals to


\begin{mathletters}
\label{allresults}
\begin{eqnarray}
        R_{\rm osc}^{\rm u}(s, x)&=&
\frac{\cos(2\pi s)}{2\pi^2y^2}P(s,x), \label{resultu}
\\
R_{\rm osc}^{\rm o}(s, x)&=&
-\frac{\cos(2\pi s)}{2\pi^4y^4}P^2(s,x),  \label{resulto}
\\
R_{\rm osc}^{\rm s}(s, x)&=&
\frac{\cos(\pi s)}{2y}P(s,x) 
-\frac{\cos(2\pi s)}{2\pi^4y^2}P^2(s,x),  
 \label{results}
\end{eqnarray}
\end{mathletters}
where $y^2=s^2+x^4$, and $P(s,x)$ is 
the spectral determinant of the diffusion operator
\begin{equation}
P(s,x)= \prod_{\mu, \epsilon_\mu 
\neq 0}\left[ \left(\frac{s}{\epsilon_\mu}\right)^2
         +\left(1+\frac{x^2}
        {\epsilon_\mu}\right)^2\right]^{-1}.
        \label{product}
\end{equation}

\noindent Note that Eq.~(\ref{Rpert}) expresses 
$R_p(s,x)$ through the Green function of this 
operator. Thus, regardless of the spectrum $\epsilon_\mu$, 
$R_p(s,x)$ and $R_{\rm osc}(s, x)$ are related:

\begin{equation}
  \label{relation}
  R_{{\rm p}}(s,x)=
  \Re\frac{1}{\alpha \pi^2(-is+ x^2)^2}- 
\frac{1}{2 \alpha \pi^2}\frac{\partial^2 \ln[P(s,x)]}{\partial s^2}.
\end{equation}

It follows from Eq.~(\ref{product}) that $R_{\rm osc}(s, x)$
together with $P(s,x)$ decays exponentially at $s\gg g$. 
As a result, the singularity in $S(\tau, 0)$ at the Heisenberg time 
 is washed out:
$S(\tau, 0)$  becomes {\it analytic} around $\tau=2\pi$.
The scale of smoothening of the singularity is 
$1/E_c$ ( see Fig.~\ref{fig:1}).
At $1\ll s\ll g$ the sum of Eq.~(\ref{allresults}) and Eq.~(\ref{Rpert}) 
gives the leading high frequency asymptotics 
of the universal result, for $s\gg g$ it coincides with the perturbative 
result $R_{{\rm P}}(s,x)$ of Ref.~\cite{Altshuler85}.

In a closed  ${\rm d}$-dimensional cubic sample  
(diffusion equation with Dirichlet boundary conditions)
$\epsilon_{\mu}=g\pi^2 \vec{n}^2$, where $\vec{n}=(n_1,\ldots, n_d)$ 
and $n_i$ are non-negative integers. 
For  $s\gg g$ and  $d<4$
  we obtain   the asymptotics             
 $P(s,0)\to \exp\{-\pi(s/\pi g)^{d/2}
/[\Gamma (d/2)d \sin(\pi d/4)]\} $. 
At $1\ll s\ll g$ we obtain 

\begin{equation}
        R(s, 0)  =   -\frac{\sin^2(\pi s)}{(\pi s)^2}+
        \frac{\sin^2(\pi s)}{\pi^2 g^2} \sum_{\vec{n}}
        \frac{1}{(\pi^2 \vec{n}^2)^2},
        \label{kkravtsov}
\end{equation}

\noindent 
This result was shown in Ref.~\cite{Kravtsov94} to be valid
 even for $s < 1$.   
Thus, it is natural to assume that for the unitary ensemble the 
sum of Eq.~(\ref{resultu}) and Eq.~(\ref{Rpert}) gives 
the correct $g \to \infty$ asymptotics  
at arbitrary frequency.  This is related to the 
absence of higher order corrections to the leading term of
the perturbation theory $S(\tau, 0)\propto \tau$ 
in the unitary case.


Now we sketch the derivation of our results.
Consider a quantum particle 
moving in a random potential  $V(\vec{r})$. 
The perturbation acting on the system is a 
change in the potential $\delta V(\vec{r})$. 
Both  $V(\vec{r})$ and $\delta V(\vec{r})$ are taken to be white noise 
random potentials with variances 
$\langle V(\vec{r})V(\vec{r'})\rangle =\delta (\vec{r}-\vec{r'})
/2\pi \nu \tau $ 
and $\langle \delta V(\vec{r})\delta V(\vec{r}')\rangle  =
x^2 \Delta \delta (\vec{r}-\vec{r'})/(4\pi \nu )$, 
$\Delta \tau\ll 1$, $\langle V(\vec{r}) \delta V(\vec{r}')\rangle  =0$,  
where $\langle \rangle $ denotes ensemble averaging and $\nu$ is 
the density of states per unit volume. 
The dimensionless perturbation strength  $x^2$
 is assumed to be of order unity.

We use the supersymmetric nonlinear $\sigma$-model 
introduced by Efetov \cite{Efetov83}, and follow his notations everywhere. 
One can show that for the  system under consideration the $\sigma$-model 
expression for $K(\omega ,x)$ is given by

\begin{mathletters}
\label{allfunc}
\begin{eqnarray}
K(\omega, x)  & = &\left. -\frac{1}{\pi^2}\Re 
\frac{\partial ^2  }{\partial J ^2}\int {\cal D} Q 
         \exp(-F_J)\right|_{J=0}
\label{kfunc}, 
\\
 F_J&=&\frac{\pi \nu}{8} \int d\vec{r} 
{\rm STr}\left\{ 
D(\nabla Q)^2 +2i \omega \Lambda  Q +i J  \Lambda k  Q- 
\frac{x^2\Delta}{2} (\Lambda  Q)^2   \right\} .  
\label{Zfunc} 
\end{eqnarray}
\end{mathletters}

\noindent The $8\times 8$ supermatrix $Q(\vec{r})$  obeys the nonlinear 
constraint $Q(\vec{r})^2=1$ and takes on its values on a certain 
symmetric space ${\bf H}={\bf G}/{\bf K}$, where ${\bf G}$ and ${\bf K}$
are groups \cite{Weidenmuller85}. For example, for the unitary ensemble 
${\bf H}={\bf SU(1,1/2)}/{\bf SU(1/1)\otimes SU(1/1)}$ \cite{Zirnbauer86}.
The integration measure for $Q$
in the functional integral  Eq.~(\ref{kfunc}) is the invariant 
measure on ${\bf H}$. 
    
The hierarchy of blocks of supermatrices is as follows: 
advanced-retarded (A-R) blocks, fermion-boson (F-B) blocks, and
blocks corresponding to time-reversal.  
$\Lambda={\rm diag}\{ 1, 1, 1, 1, -1, -1, -1, -1\}$ is the 
matrix breaking the symmetry in the advanced-retarded (A-R ) space, 
$k={\rm diag}\{1,1,-1,-1, 1,1,-1,-1\}$ is the symmetry breaking matrix 
in the Fermion-Boson ( F-B ) space. 

 The  large frequency asymptotics of 
$K(\omega,  x)$ can be obtained from Eq.~(\ref{kfunc})  
by  use of the stationary phase method. 
The conventional perturbation theory corresponds to integrating over the  
small fluctuations of $Q$ around $\Lambda$ \cite{Efetov83}, 

\begin{equation}
Q=\Lambda 
(1+iP)(1-iP)^{-1}, P= \left(\begin{array}{cc}
0 & B \\
\bar{B} & 0
\end{array}\right),
\label{parametrization}
\end{equation}

\noindent where the matrix $P$ describes these small fluctuations.

 $Q=\Lambda$ 
is not the only stationary point on ${\bf H}$. 
This fact to the best of our knowledge was not appreciated in the 
literature. 
The existence of other stationary points makes the basis for our 
main results.

It is possible to parameterize fluctuations around a point $Q_0$
in the form $Q=Q_0(1+iP_0)(1-iP_0)^{-1}$. Expanding the Free Energy 
$F_J$ in Eq.~(\ref{Zfunc}) in $P_0$ we would obtain the stationarity 
condition $\partial F_J/\partial P_0 =0$. 
This route however 
is inconvenient because the parametrization of $P_0$ will depend on 
$Q_0$. 
Instead we perform a global coordinate transformation on ${\bf H}$
that maps $Q_0$  to $\Lambda$, 
$Q_0\rightarrow T_0^{-1} Q_0 T_0=\Lambda$. 
We note that the matrices $\Lambda$ and $-\Lambda k$ 
belong to ${\bf H}$,  and the corresponding terms  
in Eq.~(\ref{Zfunc})  can be viewed as symmetry 
breaking sources. 
This transformation changes the sources, but allows us to 
keep the parametrization of Eq.~(\ref{parametrization}) 
and preserves the invariant measure. Introducing the notation 
$Q_{\Lambda}=T_0^{-1}\Lambda T_0$ and 
$Q_{\Lambda k}=T_0^{-1}\Lambda k T_0$ we rewrite Eq.~(\ref{allfunc}) as

\begin{eqnarray}
K(\omega, x)   & = & \frac{\nu ^2}{64}\Re 
\int {\cal D} Q 
\int d\vec{r} \int d\vec{r}' {\rm STr}\Big(Q_{\Lambda k}
Q(\vec{r})\Big) {\rm Str}\Big(Q_{\Lambda k} Q(\vec{r}')\Big)
\exp\left(-F(Q_{\Lambda})\right),
\nonumber  \\
F(Q_{\Lambda}) &  = & \frac{\pi \nu}{8} \int d\vec{r} 
{\rm STr}\left\{ 
D(\nabla Q)^2 +2i \omega Q_{\Lambda}  Q - \frac{x^2\Delta}{2} (
Q_{\Lambda}  Q)^2   \right\}     .
\label{kfuncsource} 
\end{eqnarray}

The stationarity condition  
$ \left. \partial F(Q_{\Lambda})/\partial P\right|_{P=0}=0 $
%
implies that  all the elements 
of $Q_{\Lambda}$ in the AR and RA blocks should vanish (this can be 
seen from Eq.~(\ref{parametrization})).

Here we  discuss in detail  only the calculation for the unitary ensemble. 
The calculation for the other cases  proceeds 
analogously, and we just  point out the important 
differences from the unitary case.


Consider now the unitary case.
The $4\times 4$ supermatrices $B$ and $\bar{B}$ 
in Eq.~(\ref{parametrization})  are  given by 
\[ B= \left(\begin{array}{cccc}
a & 0 & i\sigma _1 & 0\\
0 & a^* & 0 & -i\sigma _1^* \\
\sigma _2^* & 0 & ib & 0 \\
0 & \sigma _2 & 0 & ib^*  
\end{array}\right), 
 \bar{B}=\left(\begin{array}{cccc}
a^* & 0 & -\sigma _2 & 0\\
0 & a & 0 & \sigma _2^* \\
-i\sigma _1^* & 0 & ib^* & 0 \\
0 & -i \sigma _1 & 0 & ib  
\end{array}\right),
\]

\noindent where $a$, $b$ and their conjugates are ordinary variables, 
and $\sigma_1$, $\sigma_2$ and their conjugates are grassmann variables.

The only  matrix besides $\Lambda$ 
that satisfies the stationarity condition is 
$Q_{\Lambda}= 
\tilde{\Lambda}=-k \Lambda$. 
In this case $Q_{ \Lambda k}=  -\Lambda$.
All other matrices from ${\bf H}$ 
contain nonzero elements in the AR and RA blocks.
Both stationary points contribute substantially to 
$K(\omega, x)$.

Consider the contribution of 
$Q_{\Lambda}=\tilde{\Lambda}$ to $K(\omega, x)$ first. 
We substitute $Q_{\Lambda}=-k\Lambda$ and $Q_{ \Lambda k}=  -\Lambda$ 
into Eq.~(\ref{kfuncsource}) and expand
both the Free Energy $F(Q_\Lambda)$ and the pre-exponent to the second order 
in $B$ and $\bar{B}$. 
Expanding $B(\vec{r})$ in  the eigenfunctions of the diffusion 
operator $\phi_{\mu}(\vec{r})$ as 
$B(\vec{r})=\sum_{\mu}\phi_{\mu}(\vec{r})B_{\mu}$, and 
introducing ${\cal E}_\mu=is+\epsilon_\mu +x^2$
we arrive at the following expression for the dimensionless 
density-density correlator:

\begin{eqnarray}
        R^{\rm u}_{\rm osc}(s, x) & = & 
 \Re \int {\cal D} B_{\mu}
 \Big(\sum_{\mu}[|a_{\mu}|^2+ |b_{\mu}|^2 - 
        \sigma_{1, \mu}^* \sigma
        _{1, \mu}+\sigma_{2, \mu}^*\sigma_{2, \mu}]\Big)^2
        \nonumber  \\
         & \times & \exp( - 2\pi \{-is +\sum_{\mu}[ 
         {\cal E}_\mu|a_{\mu}|^2 +
         {\cal E}^*_\mu|b_{\mu}|^2
             - \Re{\cal E}_\mu  (\sigma_{1, \mu}^* \sigma
        _{1, \mu}-\sigma_{2, \mu}^*\sigma_{2, \mu}) 
      ] \}). 
        \label{kpt}
\end{eqnarray}

\noindent We have to keep the perturbation 
strength  $ x^2 $  finite to avoid the divergence of 
the integral over $a_0$  caused by  the 
presence of the infinitesimal imaginary part in $\omega$. 
For the non parametric case we should take the 
$x^2\to 0$ limit after the integral 
in Eq.~(\ref{kpt}) is evaluated.

Since the Free Energy in Eq.~(\ref{kpt}) 
contains no Grassmann variables in the 
zero mode they have to come from the pre-exponent. Therefore out of the 
whole square of the sum in the pre-exponent only the 
terms containing all four zero mode  Grassmann  variables
contribute. In these terms the prefactor does not contain 
any variables from non-zero modes. Thus,
the evaluation of the Gaussian integrals over non-zero modes 
yields the superdeterminant of the quadratic form in the 
exponent.
Supersymmetry around $\tilde{\Lambda}$ is broken by $s$, 
therefore this superdeterminant differs from unity and is given by
$P(s,x)$ of Eq.~(\ref{product}). 
The correctly ordered integration measure for Grassmann variables is
$\prod_{\mu} d\sigma_{1\mu} d\sigma_{1\mu}^*
d\sigma_{2\mu}^* d\sigma_{2\mu}$. 
Evaluating the integral we arrive at  Eq.~(\ref{resultu}). 

In quasi-1D for closed boundary conditions and $x=0$ 
the spectral determinant $P(s,0)$ can be evaluated exactly, 
and from Eq.~(\ref{resultu}) we obtain

\begin{equation}
        R^{\rm u, osc}_{1D}(s,0)  =  \frac{s}{2g\pi^2s^2}
        \frac{\cos( 2 
        \pi  s) }{\sinh ^2 \left(\sqrt{\frac{s}{2g}}\right)+\sin ^2 
\left(\sqrt{\frac{s}{2g}}\right)}.
        \label{k1D}
\end{equation}

For $Q_{\Lambda}=\Lambda$ the same procedure as used above 
leads to Eq.~(\ref{Rpert}), which coincides with 
the result  of Ref.~\cite{Altshuler85}.


The behavior of $S(\tau,0)$ at $\tau=0$ and $\tau=2\pi$ 
is associated respectively with $R_{\rm p}(s,0)$ (Eq.~(\ref{Rpert}))
and $R^{\rm u}_{\rm osc}(s,0)$ (Eq.~(\ref{resultu})). 
In other words the singularity 
at the Heisenberg time is determined by the contribution 
to $R(s,0)$ from ${\tilde{\Lambda}}$.  
It is clear that the cusp in $S(\tau, 0)$ at $\tau =2\pi $
will be rounded off  because $R^{\rm u}_{\rm osc}(s,0)$ 
decays exponentially at large $s$. The scale of the smoothening 
is of order $1/g$.   

The Fourier transform of Eq.~(\ref{k1D}) ( see Fig.~\ref{fig:1} ) is 
\begin{equation}
    S^{\rm u}_{1D}(2\pi +t, 0)_{\tilde{\Lambda}}  = 
 \sum_{n=1}^{\infty}\frac{(-1)^n \exp(-\pi^2 n^2 g |t|)}{\pi^2 g 
 n \sinh(\pi n)}-\frac{|t|}{4\pi} .
        \label{ktildefourier}
\end{equation}

\noindent Even though $S^{\rm u}_{1D}(2\pi +t, 0)_{\tilde{\Lambda}}  $
appears to be  a function of $|t|$, it is  regular at $t=0$. 

We can also estimate 
$S^{\rm u}(2\pi , 0)_{\tilde{\Lambda}}$
in any dimension. It is proportional to $1/g$ of Eq.~(\ref{gdef}) 
and is given by 

\[ S^{\rm u}(2\pi , 0)_{\tilde{\Lambda}} 
= \frac{1}{4 \pi^4 g} \int_{-\infty+i\eta}^{\infty+i\eta} \frac{d z}{ z^2} 
\prod _{\mu \epsilon_\mu \neq 0}\left( 1+\left[\frac{z\epsilon_1}
{\epsilon_\mu}\right]^2\right)^{-1} .\]



Consider now T-invariant systems.
For the orthogonal ensemble there are still only two stationary points 
on ${\bf H}$: $\Lambda$ and $\tilde{\Lambda}= -k \Lambda$. 
To determine the contribution of the $\tilde{\Lambda}$-point 
we  use the formula Eq.~(\ref{kfuncsource}) 
 with   $Q_{\Lambda}=\tilde{\Lambda}$ and $Q_{k \Lambda }=-\Lambda$
and Efetov's  parametrization for the perturbation theory \cite{Efetov83}.
The calculations are analogous to those for the 
unitary ensemble and lead to  Eq.~(\ref{resulto}).
The contribution of $Q_\Lambda=\Lambda$ gives Eq.~(\ref{Rpert}).
At  $\tau=2\pi$ the third derivative of $S(\tau,0)$  for the orthogonal 
ensemble has a jump. This 
singularity also disappears at finite $g$. 


In the symplectic case there are three types of stationary 
points which correspond to singularities in the 
structure factor $S(\tau,0)$ at $\tau=0, \pi , 2\pi$  \cite{Mehta91}. 
The $\tau=2\pi$ singularity corresponds to $Q_{\Lambda}=\tilde{\Lambda}=
-k\Lambda$, and 
its contribution to $R(s, x)$,
given by the second term in Eq.~(\ref{results}), is 
exactly the same as $R^{\rm o}_{\rm osc}(s,x)$.
The stationary point $Q_{\Lambda}=\Lambda$ corresponds
to the $\tau=0$ singularity in $S(\tau,0)$ and leads to Eq.~(\ref{Rpert}). 
The $\tau=\pi$ singularity corresponds to a degenerate manifold 
of matrices $Q_{\Lambda}$ on ${\bf H}$ 
$Q_\Lambda={\rm diag}(\tau_{\vec{m}},\openone_2, -\tau_{\vec{m}},
-\openone_2)$,  
$ Q_{k\Lambda} =-k Q_\Lambda $,
where ${\openone_2}$ is a $2\times 2$ unit matrix, 
$\tau_{\vec{m}}=m_x \tau_{x}+m_y\tau_{y}$, 
$\vec{m}^2=1$  and $\tau_{x,y}$ are Pauli matrices in the time-reversal 
block. 
%
%
The calculation proceeds as before and leads to the first term in 
Eq.~(\ref{results}).
In quasi-1D we can obtain the leading contribution to the structure factor 
$S(\tau, 0)$ around $\tau=\pi$ 
\begin{equation}
  \label{formfsympl}
  S^{\rm s}(t+\pi,0)=\int_0^{\infty} 
\frac{-4\sin^2(g|t|z)dz}{\sinh^2\sqrt{z}+\sin^2\sqrt{z}}+\ln(1.9 g)+ O(1/g).
\end{equation}

\begin{figure}
  \begin{center}
  \epsfxsize=15cm \epsfbox{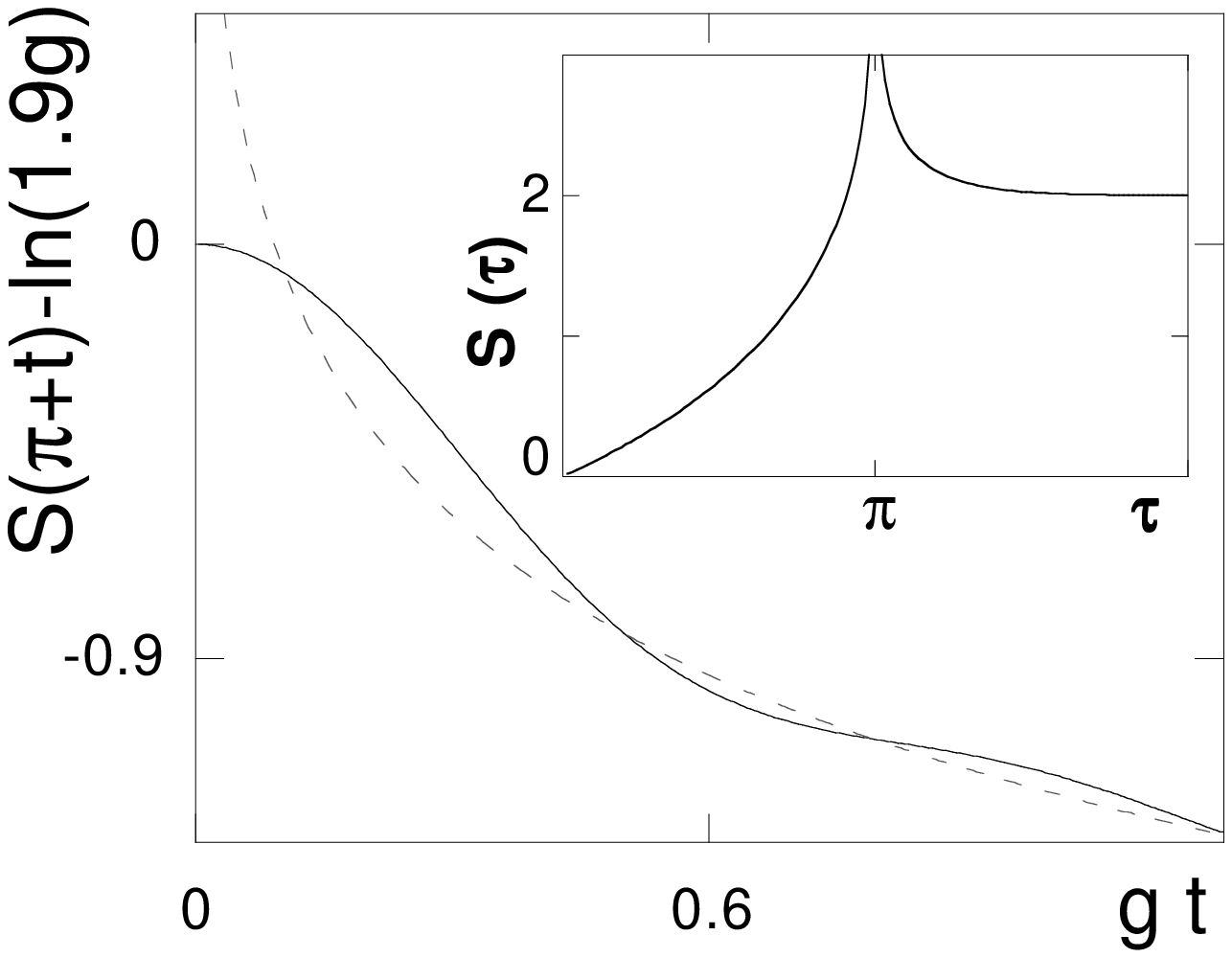}
  \end{center}
  \caption{The  structure factor for the 
    symplectic case in quasi-1D (solid line) and 
    the universal result (dashed line). 
    Insert: the universal stucture factor.}
  \label{fig:2}
\end{figure}

\noindent  
The result is plotted in Fig.~\ref{fig:2}. In all dimensions the 
 logarithmic divergence 
in the zero mode result is now cut off by finite $g$, and 
$S^{\rm s}(\pi,0)\propto \ln g$.

In conclusion we mention several points about our results. 
1) Equation (\ref{allresults}) describes the deviation of  
the level statistics of a weakly 
disordered chaotic grain  from the universal ones. 
This deviation is controlled by the diffusion operator. This
operator is purely classical. 
It seems plausible that the nonuniversal part of spectral statistics 
of any chaotic system can be expressed through a spectral determinant 
of some classical system-specific operator. If so, the relation 
Eq.~(\ref{relation}) should be universally correct!
%
%
%
 
2) The formalism used here should be 
applicable even to the systems 
weakly coupled to the outside world (say through tunnel contacts). 
As long as the level broadening $\Gamma$ ($\omega=\Re\omega +i\Gamma$) 
is smaller than $\Delta x^2$ 
 the integration over the zero mode variables 
in Eq.~(\ref{kpt}) is convergent. The integral over the other modes 
is always convergent provided $\Gamma < E_c$. 
Thus, the presence of a perturbation can effectively ``close'' a 
weakly coupled system. Under these conditions Eq.~(\ref{allresults}) 
remains valid after the substitution $\cos(2\pi s)\to 
\exp(-2\pi\Gamma/\Delta)\cos(2\pi s)$ and $x^2\to x^2-\Gamma/\Delta$.

3)
The classification of physical systems into the three universality 
classes (unitary, orthogonal and symplectic) is, of course, an 
oversimplification. In practice there is always 
a time scale which determines the crossover from one 
ensemble to another. For example if a system is subjected to 
a magnetic field for very short times it 
will still effectively remain orthogonal. 
On the other hand, the long time behavior will be unitary. 
The characteristic time is 
set by the strength of the magnetic field.

For a disordered metallic grain in a magnetic field this 
characteristic time is given by 
$l^2_{H}/D$. For  frequencies larger than  $D/l^2_{H}$ 
the system effectively becomes orthogonal. 
This implies that even if we neglect the spatially nonuniform 
fluctuations of the $Q$-matrix the cusp in $S(\tau,0)$ at $\tau=2\pi$ 
will be washed out on the scale of $\Delta l_H^2/D$
( although there will still remain a jump in the third derivative of 
$ S(\tau, 0)$ ). For the system to behave as  unitary for frequencies 
of order $E_c$  the magnetic length $l_H$ has to be 
shorter than the size of the 
system. Spin-orbit interaction that causes the orthogonal-to-symplectic 
crossover can be considered in a similar way. 

4) The rounding off of the singularity in $S(2\pi,0)$ is also 
present in the random matrix model with preferred basis \cite{Pichard93} 
\cite{Shapiro94}. Note that our results differ from those in 
Ref.~\cite{Shapiro94} substantially. This means that finite $g$ is not 
equivalent to finite temperature for the corresponding 
Calogero-Sutherland model \cite{Simons93}.

We are  grateful to D.~E.~Khmel'nitskii, B.~D.~Simons and N.~Taniguchi 
for numerous discussions 
throughout the course of this work.

\begin{references}

\bibitem{LesHouches89} Chaos in Quantum Physics, 
eds., M.~J.~Jianonni, A.~Voros and  J.~Zinn-Justin, Les Houches,
Session LII 1989 (North-Holland, Amsterdam, 1991) 

\bibitem{Thouless} D.~J.~Thouless, Phys. Rep. {\bf 13}, 93 (1974) 

\bibitem{Mehta91} M.~L.~Mehta, Random Matrices, (Academic Press, 
New York, 1991)

\bibitem{Dyson} F.~J.~Dyson, J. Math. Phys. {\bf 3}, 140, 157, 166 
(1962) 

\bibitem{Haake}F.~Haake, Quantum Signatures of Chaos (Springer, Berlin, 
1991)

\bibitem{Prigodin94} The function $S(\tau,0)$, for example, describes  
the phenomenon of ``quantum echo'' in mesoscopic quantum dots. 
See V.~N.~Prigodin et al, Phys. Rev. Lett. {\bf 72}, 546 (1994)

\bibitem{Berry85}M.~V.~Berry, Proc. Roy. Soc. London {\bf A 400}, 
229 (1985)

\bibitem{Kravtsov94}V.~E.~Kravtsov, A.~D.~Mirlin, Sov. Phys. JETP Lett.,
{\bf 60}, 656 (1994). [Pis'ma ZhETF, {\bf 60}, 645 (1994) ]

\bibitem{Altshuler85}B.~L.~Altshuler, B.~I.~Shklovskii, JETP 
{\bf 64}, 127 (1986).


\bibitem{footnote}This definition of $\alpha$ is related to the fact that 
 $\Delta$ for the unitary and orthogonal cases 
is the mean level spacing per spin polarization, and for the 
symplectic case it is one-half of that. 

\bibitem{Efetov83} K.~B.~Efetov Adv. Phys. {\bf 32}, 53 (1983).

\bibitem{Weidenmuller85} J.~J.~Verbaarscot, H.~A.~Weidenmuller, 
M.~R.~Zirnbauer, Phys. Rep. {\bf 129}, 367 (1985)

\bibitem{Zirnbauer86} M.~R.~Zirnbauer, Nucl. Phys.{\bf B [FS] 265},
375 (1986)

\bibitem{Pichard93} J.~-L.~Pichard, B.~Shapiro J. Phys. May(1994)

\bibitem{Shapiro94}M.~Moshe, H.~Neunberger, B.~Shapiro, 
preprint RU-94-28, Technion PH-12-94 (1994)

\bibitem{Simons93} B.~D.~Simons, P.~A.~Lee and B.~L.~Altshuler, 
Nucl. Phys.{\bf B 409}, 487 (1993)


\end{references}
\end{document}